\documentclass[modern]{aastex61}

\usepackage{graphicx}
\usepackage{amsmath}
\usepackage{amssymb}
\usepackage{enumitem}

\newcommand\mybar{\kern1pt\rule[-\dp\strutbox]{.8pt}{\baselineskip}\kern1pt}

\setlist[itemize]{noitemsep, topsep=0pt, leftmargin=*}

\shorttitle{SIDM from Gravitational Scattering}
\shortauthors{Loeb}

\begin{document}

\title{Effective Self-Interaction of Dark Matter from Gravitational Scattering}

\author{Abraham Loeb}
\affiliation{Astronomy Department, Harvard University, 60 Garden
  St., Cambridge, MA 02138, USA}

\begin{abstract}
I show that gravitational scattering of dark-matter objects of mass
$\sim 10^4M_\odot$ and speeds of $\sim 10~{\rm km~s^{-1}}$, provides
the cross-section per unit mass required in self-interacting dark
matter models that alleviate the small-scale structure challenges to
the collisionless cold dark matter model. For primordial objects of
mass $10^4M_4~M_\odot$, moving at the velocity dispersion
characteristic of dwarf galaxies, $10v_1~{\rm km~s^{-1}}$, the
cross-section per unit mass for gravitational scattering is $\sim
10~(M_4/v_1^4)~{\rm cm^2~g^{-1}}$. The steep decline in interaction
with increasing velocity explains why self-interaction is not evident
in data on massive galaxies and clusters of galaxies.
\end{abstract}

\section{Introduction}

Self-interacting dark matter (SIDM)
\citep{2000PhRvL..84.3760S,2000MNRAS.315L..29F} could solve the
small-scale structure challenges to the standard cosmological model of
cold dark matter \citep{2016PhRvL.116d1302K,2017ARA&A..55..343B}. Many
studies over the past two decades demonstrated that a self-interaction
cross section per unit mass, $\sigma/m$, in the range $(1$--$10)~{\rm
  cm^2~g^{-1}}$ modifies the expected dark matter cusps to central
cores - as suggested by observations of dwarf galaxies
\citep{2001ApJ...547..574D,2002ApJ...581..777C,2012MNRAS.423.3740V,2013MNRAS.430...81R,2013MNRAS.431L..20Z,2014MNRAS.444.3684V,2015MNRAS.452.1468F,2017PhRvL.119k1102K,2017MNRAS.468.2283C,2017MNRAS.472.2945R,2018MNRAS.479..359S,2018PhR...730....1T,2019MNRAS.490..962F,2020PhRvL.124n1102S,2022arXiv220306035M,2022arXiv220310104S},
and resolves the ``too-big-to-fail'' challenge
\citep{2012MNRAS.423.3740V,2019MNRAS.490..231K,2021MNRAS.505.5327T}.
We note, however, that the uncertainty in modeling the baryonic
component is still large enough to also offer a possible solution to
these small-scale challenges \citep{2017ARA&A..55..343B}.

Since there is no apparent discrepancy with cold dark matter on the
scales of massive galaxies or groups (with $v\gtrsim 10^2~{\rm
  km~s^{-1}}$) and clusters of galaxies (with $v\sim 10^3~{\rm
  km~s^{-1}}$), \citet{2011PhRvL.106q1302L} proposed a decade ago that
the interaction might be mediated by a Yukawa potential, declining
inversely with velocity to the fourth power, $\propto v^{-4}$, as
expected in some dark sector extensions to the standard model of
particle physics (see also, \citet{2021JHEP...06..008C}). Most
recently, velocity-dependent cross-sections with values $\gtrsim
5~{\rm cm^2~g^{-1}}$ at $v\lesssim 10~{\rm km~s^{-1}}$ were motivated
to explain the dynamical properties of Milky-Way satellites
\citep{2022arXiv220310104S}, but with the provision that the
interaction must drop sharply with velocity to $\ll 1~{\rm
  cm^2~g^{-1}}$ in massive systems
\citep{2016PhRvL.116d1302K,2022arXiv220212247S}.

Here we point out that the normalization and velocity dependence of
the cross-section per unit mass required to alleviate the small-scale
structure challenges to the cold dark matter model, is provided
naturally by gravitational scattering if the dark matter is composed
of objects in the mass range of $10^{3}$--$10^{4}M_\odot$ and a
physical size $\lesssim 1~{\rm pc}$. The considerations are presented
in the next section and the implications are summarized in the
concluding section.

\section{Cross-Section for Gravitational Scattering}

The gravitational cross-section for scattering of compact
objects with mass $m$ and characteristic velocity $v$ is given by
\citep{1962pfig.book.....S,1987gady.book.....B},
\begin{equation}
\sigma = 8\pi\times \left({Gm\over v^2}\right)^2 \ln\Lambda ,
\label{one}
\end{equation}
where $\ln \Lambda=\ln(b_{\rm max}/b_{\rm min})$ is the Coulomb
Logarithm, determined by the ratio between the maximum and minimum
values of the impact parameter, with $b_{\rm max}\sim
(4\pi\rho/3m)^{-1/3}$ being the average separation between objects at
a mass density $\rho$, and $b_{\rm min}\sim 2Gm/v^2$ is the impact
parameter for a 90-degree deflection. The characteristic parameters in
the cores of dwarf galaxies, $v \sim 10~{\rm km~s^{-1}}$ and $\rho
\sim 3\times 10^7~M_\odot~{\rm kpc^{-3}}$, yield, $\ln \Lambda \sim 4$
for $m\sim 10^4M_\odot$.

Dividing the cross-section by the object's mass, we get, 
\begin{equation}
{\sigma \over m} = 10~{\rm cm^2~g^{-1}}~\left[{(m/10^4M_\odot)\over
    (v/10~{\rm km~s^{-1}})^4}\right] .
\label{two}
\end{equation}
Remarkably, for the mass range of $m \sim (10^3$-$10^4)M_\odot$,
gravitational scattering provides the normalization and velocity
dependence required for alleviating the small-scale structure challenges
to the cold dark matter model.

The physical size of the dark matter object, $R$, must be smaller than
the minimum impact parameter for their gravitational scattering,
\begin{equation}
R_{\rm max} = \left({2Gm\over v^2}\right)= 1~{\rm pc} \left[{(m/10^4M_\odot)\over
    (v/10~{\rm km~s^{-1}})^2}\right],  
\label{three}
\end{equation}
implying a mass density inside each object that exceeds the value,
\begin{equation}
\rho_{\rm min} = \left[{m\over (4\pi/3) R_{\rm max}^3}\right]= 2\times
10^{12}~M_\odot~{\rm kpc^{-3}} \left[{(v/10~{\rm km~s^{-1}})^6\over
    (m/10^4M_\odot)^2}\right].
\label{four}
\end{equation}

This minimum density of a characteristic value, $\sim 2 \times
10^{-19}~{\rm g~cm^{-3}}$, is $7\times 10^{10}$ larger than the mean
cosmic density of matter in the present-day universe and corresponds
to a minimum formation redshift of $\gtrsim 700$, around the cosmic
epoch of hydrogen recombination. The origin of such dark matter
objects must therefore be primordial since the standard power-spectrum
of density fluctuations forms the first virialized mini-halos at
redshifts $z\sim 70$
\citep{2010hdfs.book.....L,2013fgu..book.....L,2014IJAsB..13..337L}.

\section{Implications} 

We have shown that gravitational scattering of compact objects could
provide the cross-section per unit mass required in self-interacting
dark matter models that alleviate the small-scale structure challenges
to the collisionless cold dark matter model. For primordial objects of
mass $10^3$--$10^4~M_\odot$, the cross-section for gravitational
scattering is $\sim 1$--$10~{\rm cm^2/g}$, at the velocity dispersion
characteristic of dwarf galaxies, $\sim 10~{\rm km~s^{-1}}$. The sharp
decline in the cross-section at higher velocities, $\propto v^{-4}$,
explains why self-interaction is not evident in data on massive
galaxies or clusters of galaxies \citep{2016PhRvL.116d1302K}. Much
larger values of the cross-section, corresponding to a higher mass
$m$, are disfavored since they trigger gravothermal core collapse
\citep{2021MNRAS.505.5327T}.

Ultra-faint galaxies, such as Segue 1 and 2
\citep{2009ApJ...704.1274W} possess velocity dispersions of a few
${\rm km~s^{-1}}$ where scattering should be more pronounced. They
offer excellent laboratories for testing the model proposed
here. Additional constraints on the existence of massive dark-matter
objects can be derived from the comparison between data (from Gaia,
HST, JWST and LSST) and numerical simulations of cold streams in a
clumpy Milky Way halo
\citep{2020ApJ...892L..37B,2021hst..prop16791B,2021MNRAS.502.2364B,2021MNRAS.504..648B}.

Primordial black holes (PBHs) in the required mass range are
constrained by microlensing of supernovae and of stars, as well as by
wide binaries and X-ray binaries; for a compilation of all related
limits, see Figure 1 in \citet{2021arXiv211002821C}. The
characteristic value of $R_{\max}$ in equation (\ref{three}) is larger
than the Einstein radius of microlenses or the typical separation of
wide binaries; in addition, extended objects need not trigger
substantial X-ray luminosity from accretion of baryons.  Therefore,
the above PBH constraints might be relaxed for objects that are not as
compact as black holes. In order to resolve the small-scale structure
challenges, the objects under consideration here must make most of the
dark matter. 

If the required objects resulted from a cosmological phase transition
at a temperature $T$, then their mass is expected to reflect the
horizon mass,
\begin{equation}
m_{\rm H} \sim 10^4M_\odot\left({T\over 2~{\rm MeV}}\right)^{-2} .
\label{four}
\end{equation}
Interestingly, the required mass range is naturally realized during
the weak-interaction epoch, after the QCD phase transition at $T\sim
200~{\rm MeV}$ and before neutrino decoupling at $\sim 1~{\rm MeV}$.
This is well above the minimum redshift for the production of cold
dark matter \citep{2015JCAP...03..004S}.

Another possible origin of the dark matter clumps is that they were
the first objects to collapse gravitationally after cosmological
recombination (in the redshift range of $z\sim 10^2-10^3$) as a result
of large density fluctuations on their mass scale.  In this case, the
dark matter can still be collisionless at the elementary particle
level. These objects would evade microlensing and wide-binaries
constraints on PBHs because they are extended and fluffy.  The
standard cosmological model makes the first collapsed objects at
redshift $z\sim 70$
\citep{2001PhR...349..125B,2007MNRAS.377..667N,2013fgu..book.....L,2016arXiv160608926L}. Interestingly,
the baryonic Jeans mass is $\sim 10^4M_\odot$ at that redshift
\citep{2016arXiv160608926L}, but larger than standard primordial
fluctuations are needed to clump most of the dark matter into the
objects of sub-parsec size discussed here.
\bigskip
\bigskip
\section*{Acknowledgements}

This work was supported in part by Harvard's {\it Black Hole
  Initiative}, which is funded by grants from JFT and GBMF. I thank
Sunny Vagnozzi for helpful comments on the final manuscript.

\bigskip
\bigskip
\bigskip

\bibliographystyle{aasjournal}
\bibliography{d}
\label{lastpage}
\end{document}